# Detecting Botnet Activities Based on Abnormal DNS traffic

Ahmed M. Manasrah, Awsan Hasan
National Advanced IPv6 Center of Excellence
Universiti Sains Malaysia,
Pulau Pinang, Malaysia
{ahmad,awsan}@nav6.org

Omar Amer Abouabdalla, Sureswaran Ramadass
National Advanced IPv6 Center of Excellence
Universiti Sains Malaysia,
Pulau Pinang, Malaysia
{omar, sures}@nav6.org

**Abstract— The botnet is considered as a critical issue of the Internet due to its fast growing mechanism and affect. Recently, Botnets have utilized the DNS and query DNS server just like any legitimate hosts. In this case, it is difficult to distinguish between the legitimate DNS traffic and illegitimate DNS traffic. It is important to build a suitable solution for botnet detection in the DNS traffic and consequently protect the network from the malicious Botnets activities. In this paper, a simple mechanism is proposed to monitors the DNS traffic and detects the abnormal DNS traffic issued by the botnet based on the fact that botnets appear as a group of hosts periodically. The proposed mechanism is also able to classify the DNS traffic requested by group of hosts (group behavior) and single hosts (individual behavior), consequently detect the abnormal domain name issued by the malicious Botnets. Finally, the experimental results proved that the proposed mechanism is robust and able to classify DNS traffic, and efficiently detects the botnet activity with average detection rate of 89%.**

## I.    INTRODUCTION

The growth in the area of network in the past few years is considered as a part of the exponential growth of the communication system. The network is just like computers; it needs software to simplify its functionality and makes it easy to use. Internet browsing, e-mail, and instant messaging are a few examples of the usage of computer communication over the Internet. Nowadays, personal computer systems are widely used, hence the number of Internet subscribers have increased gradually. Generally, these computers contain important data, such as users' information and probably any business activities [9].    Therefore, the computers have become a favorite target that attracts the attacker s' community. Even though, these systems are protected by antivirus software and firewalls, they may still be exposed to different malicious attacks. Especially, those attackers are always looking for various techniques to assist them in compromising a large number of computer systems in the world [2].

Nowadays, Botnet is considered as a serious problem as it forms a major and dangerous part of the Internet. This is because it spreads rapidly in the network over the Internet, and it is difficult to be detected because they have the ability to hide themselves as the virus and propagate as the network worms [9, 21].

## II.    BOTNET PHENOMENON

Botnet consists of a collection of Bots running on a compromised computer, which can be remotely controlled by an attacker called "botmaster" via the command-and-control (C&C) server.  Importantly, these Bots are individual piece of programmable software. It can be installed and run automatically in any compromised system, and it has the ability to spread similar to the worms', and also it can evade any detection programs similar to viruses [26]. So any compromised network infected with a large number of Bots, is called a botnet [1].

All Bots receive and execute the same command from the botmaster and respond to the same C&C server for an execution result [5]. The Botnet listens to a particular channel (i.e. IRC and HTTP) in C&C server [14] to receive further instructions from the botmaster [7]. These channels are used to carry out commands issued by the botmaster to the Bots [18]. In most cases, the C&C server is a compromised system under the control of the botmaster who is controlling the entire Botnet. The Bots need to communicate with the C&C server regularly to receive more instructions from the botmaster [5]. Therefore, if the network administrators or authorities block the C&C server, the Bots cannot receive the commands issued by the botmaster. In this case, the botmaster will compromise a new C&C server and use the Dynamic Domain Name System (DDNS) to move his domain name from the old C&C server to the new C&C server [18]

The DNS is a distributed database spread over the Internet, which is used to translate the domain names into IP addresses and vice versa [14,23]. Thus, by using the DNS,

This work was supported by the Research University Grant 1001/PPTMK/817022, Distributed Network Monitoring and Security Platform, Universiti Sains Malaysia,

"botmaster" could direct the Bots to migrate to the domain name, which has been moved to new C&C server as the domain name is hard-coded in the Bots' binary. The botnet queries the DNS server to find out the "botmaster's" domain name. In return, the DNS server replies to the Bots and provides them with the new IP address of the "botmaster's" domain name, which is located in a new compromised C&C server [18]. Nowadays, the DNS has become the desired target of "botmasters" due to its importance in the Internet. DNS is not owned or controlled by a specific organization and the DNS traffic flows between the clients and DNS server without any protection or restriction. As such, the Botnet can exploit the DNS to perform their malicious activities. The Botnet queries the DNS server just like any legitimate host and the DNS server responds to this query without distinguishing the source of the query [3].

### III. BOTNET AND DNS

There are many computer applications and legitimate users who utilize the DNS to access the Internet and perform their jobs correctly [24]. On the other hand, Botnet also utilizes the DNS to perform its malicious activities. Since many normal applications require DNS to access the Internet, the problem persists in how the normal DNS traffic caused by a legitimate user or application can be distinguished from the abnormal DNS traffic caused by the Botnet activity. However, by monitoring the DNS traffic, it is possible to identify and detect the Botnet in DNS traffic [11,24].

### IV. BOTNET BEHAVIOR

Akiyama *et al.* (2007) proposed three important behaviors of botnet, which was discovered by monitoring the activities of Botnets during the flow of data in the C&C servers [1]. These behaviors are:

- **Bots Relationship:** The relationship between botmaster and Bots is one to many, because the botmaster usually controls a number of Bots and issues the same command to all the Bots. Hence, the Bots work as one group and it is possible to detect their behavior by monitoring the activities of these groups of Botnet in the network traffic.

- **Bots Synchronization:** Botnet receives the same command from the botmaster. They communicate between each other and attack at the same time. This action can expose the group of Botnet, because the ratio of traffic that is released from this group is very high compared to the others and in some cases these traffic are discrete in time.

- **Bots Responding:** When the Bot receives commands from the botmaster, it responds immediately to those commands and executes them accurately. When the Bot receives a command from the botmaster, it executes it immediately without a need to think about it, so the time taken to do this is always constant. Thus, this response time can be used to discover the presence of the Botnet.

Monitoring the botnet behaviors and exploit it, is considered as one of the detection keys of the Botnet activities in DNS traffic. Choi *et al.* (2007) discussed some Botnet features in the DNS, and how the Botnet could exploit the DDNS to move to new C&C server when the old one is blocked. If so, the Botnet queries the DNS server to find the location of the domain name [4]. Table 1 shows the comparison between the activities of a legitimate host and Botnet when both are using the DNS.

TABLE I.     DIFFERENCE BETWEEN BOTNET AND LEGITIMATE HOSTS

| Using DNS By | Requested Domain Name | Activity and Appeared Pattern |
|---|---|---|
| Botnet | *Botnet members have fixed group size* | *Group appears immediately* |
| Legitimate Host | *Anonymous legitimate users have random size* | *Usually appears randomly and continuously* |

### V. DNS MONITORING

There are several researches conducted with regards to this problem, these researches focused on distinguishing between the normal DNS traffic generated legally in the monitored network, and those suspicious and alike to Botnet behavior.

Kristoff, (2004) conducted a study that monitors the DNS, in order to detect the botnet with prior knowledge of the blacklisted servers that spread or connect to malicious malware. This approach can simply evade when the botmaster knows this mechanism, hence it could be easily tricked by using fake DNS queries [11]. Therefore, Weimer, (2005) conducted another study to monitor the DNS traffic. The study was in passive DNS replication. The purpose of it was to build a reverse lookup with IP addresses for which no PTR records exist. By doing so, it will be easy to detect any domain name used to contact a system on the Internet [23].

Dagon (2005) discovered that the ratio of abnormal DNS traffic is high compared to the others and this indicates the presence of botnet activity. But this approach generates false results and could classify the legitimate domain name as abnormal domain name [6].

Ramachandran *et al.*'s (2006) proposed a technique and heuristics by utilizing the DNSBL blacklist lookup traffic to identify the botnet, where the technique performs counter-intelligence that detects DNSBL inspection on the botnet activity group that spreads the mail spam. But this technique also generates false positives due to the active nature of counter-measures such as inspection poisoning. Besides, this approach could not detect the distributed inspection [16]. As a result, Schonewille and Van Helmond (2006) proposed an approach based on the abnormal frequency of NXDOMAIN reply rates. However, the approach could detect several abnormal domain names effectively and generate less false positives [20]. Choi *et al.*'s (2007) study detects the botnet by exploiting the group activity feature of the botnet. This approach is stronger than the previous approaches but the main weakness of this approach is when it is applied to large scale network as the processing time will be higher [4]. Finally Tu *et al.* (2007) conducted a study to identify the activities of botnets by mining the DNS traffic data [22].

Meanwhile, the proposed approach in this paper does not require any prior knowledge of blacklisted servers to classify the DNS. Besides, it also does not depend on the high ratio of DNS traffic to detect the botnet. However, it depends on exploitation of the Botnet's behavior in the DNS traffic, particularly the appearance of botnet as a periodic group of hosts. The probability of botnet detection can be obtained by measuring the ratio of similarities between any blocks of the hosts that requested the same domain name at any given time interval.

## VI. The Proposed Method

The proposed mechanism refers to monitoring and capturing the DNS traffic at different time intervals $t_i$, and measure the ratio of similarity between any two blocks of hosts $X$ and $Y$ (group behavior) requesting the same domain name at time intervals $t_{i1}$ and $t_{i2}$. Therefore, the Jaccard similarity coefficient $S_j$ is chosen because it is simple and provides good results [4, 17]. Jaccard similarity coefficients consist of three summation variables: $X$, $Y$ and $Z$ as shown in Equation 1:

$$S_j = \frac{z}{z + x + y} \qquad (1)$$

$Z$ is the number of similar elements that are in both two objects $X$ and $Y$.

$X$ is the number of elements in the first object $X$ only but not in $Y$.

$Y$ is the number of elements in the second object $Y$ only but not in $X$.

Table 2 clarifies the probability value of the Jaccard similarity coefficient $S_j$ when used to match between two blocks of hosts.

TABLE II.    Jaccard Similarity Values

| Jaccard Similarity Value | Probability |
|---|---|
| $S_j = 1$ | Similarity ratio between all the hosts in the two blocks is 100%. So this domain name is an abnormal domain name issued by Botnet activity. |
| $S_j \geq 0.8$ And $S_j \leq 1$ | There is an assurance that 80% of the hosts make association in a direct or indirect relation. It is a good value for this research (considering false alarm rates and network delay time). Hence, this domain name is an abnormal domain name issued by Botnet activity. |
| $S_j \geq 0$ And $S_j < 0.8$ | The similarity ratio is less than 80% as it cannot be stated exactly that there is a similarity between the two blocks of hosts. So this domain can classify as normal domain name. |
| $S_j = 0$ | There is no similarity between hosts in the two blocks and consequently the domain name is normal domain name. |

To apply the Jaccard similarity values between the two blocks of hosts, the MAC address is a preferred choice as the host's identifier rather than the IP address. This is because the botmaster exploits the feature of dynamic IP that may hide the identity of the infected hosts with the Bot. Consequently; it is not reliable to place this IP on the blacklist. The Dynamic Host Configuration Protocol (DHCP) assigns multiple dynamic IP addresses to the unique host. Any infected host such as laptops can move from one network to another with new IP address assigned to it each time it connects to a new network. This forms the host's identifier by tracing the IP address aliasing and generating false information about the activity of this host [25].

By using the MAC as host's identifier to identify the activity of the hosts, accurate results could be obtained even if the hosts move from one location to another, because the DHCP cannot act on the MAC address. Moreover, the infected hosts that caused this abnormal traffic can be detected. However, the MAC address spoofing is not taken into consideration, because any Bot infected host would want the reply back to itself when sending a query to the DNS server. In the case of spoofing, the reply is sent back to different host which is out of the scope of this research. However, the spoofing takes place in another scenarios such as DDoS attack.

## VII. MONITORING NORMAL AND ABNORMAL DNS TRAFFIC BEHAVIORS

The ratio of abnormal traffic in the C&C server appeared to be higher compared to the normal DNS traffic in the case of Botnet [6]. This abnormal DNS traffic appears only in a short and discrete time, but the activity of a legitimate host appears for a longer and maybe continuous time as shown in Figure 1.

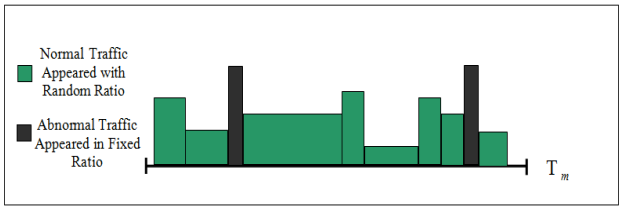

Figure 1. Normal and Abnormal DNS Traffic

The "botmaster" instructs all the Bots to perform their malicious activities simultaneously as groups in a short and discrete time and then stop all these activities suddenly, and so on. Taking this important behavior into consideration, the detection of Botnet can be made possible.

This method relies on monitoring the DNS traffic for certain time $T_m$; this time is divided into different time intervals $t^{i1}$ and $t^{i2}$. A relationship is formed between any two blocks of hosts requesting the same domain name and calculates the probability of botnet detection between these two blocks of hosts by using Jaccard similarity as portrayed in Figure 2. The probability of botnet detection $P_{bots}$ is possible if the size of block X and Y is not equal to zero and if the DNS ratio $R^{DNS}$ within the monitoring time $T_m$ is also greater then zero as depicted in Equation 2:

$$P_{bots} \begin{cases} \|X\| and \|Y\| \neq 0 \\ \\ R_{DNS} \text{ in } T_m > 0 \end{cases} \qquad (2)$$

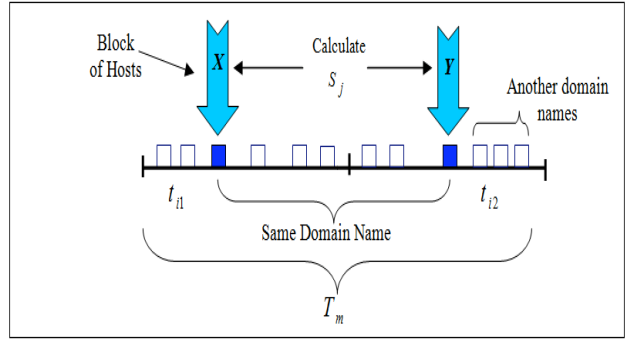

Figure 2. Applying Jaccard Similarity between Two Blocks of Hosts

A simple mechanism framework is created to classify the DNS traffic and detect the Botnet activity in DNS; it is called the Botnet Detection Mechanism (BDM). The BDM consists of three main phases: capturing phase, analyzing phase, and classifying phase as illustrated in Figure 3.

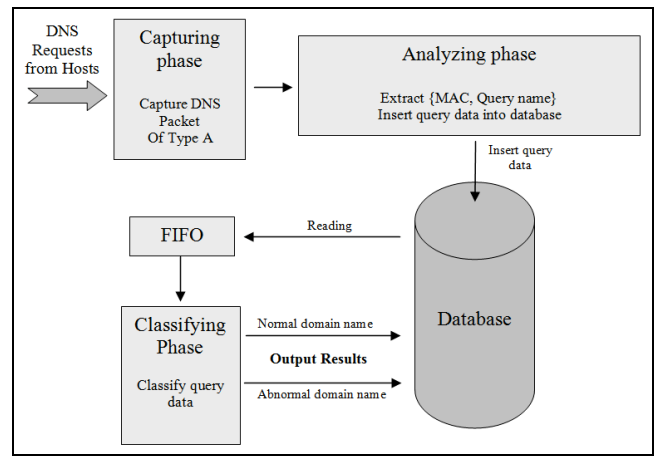

Figure 3. BDM Framework

The BDM counts the query data at each time intervals for any domain name requested by blocks of hosts during the monitoring time $T_m$. If there are two groups/blocks of hosts requesting the same domain name at time interval $t_{i1}$ and $t_{i2}$, then the BDM applies the Jaccard similarity $S_j$ between these groups of hosts and uses the MAC addresses as the host's identifier. The BDM performs this by measuring the ratio of overlapping MAC addresses in the two groups. Hence, the BDM stores the results in the database as normal domain name issued by legitimate hosts or abnormal domain name issued by botnet activity as per the similarity probabilities detailed in Table 2. If an abnormal domain name is found, the BDM sends alarm to the network administrator to block this abnormal domain name. The MAC addresses of these infect hosts are marked as blacklist.

Since bots queries the DNS server as a group of hosts periodically, If there is any domain name requested by a single host at different time intervals then it could be a normal domain. Therefore, the probability is calculated for this single host *(could be an infected host used by the botmaster to check his domain validity in C&C server)*. This host activity usually occurs before the block of infected hosts queries the domain name that has been checked by the "botmaster". The BDM performs checking at every single host on whether it is a bot infected host or legitimate host. This can be done by matching the domain name requested by this single host with the normal and abnormal domain name stored in BDM database and requested by groups of hosts.

If a single host that had requested a normal domain name which is stored in the database then, it cannot be clearly stated that this is a normal domain name because the infected host also requests for normal domain name due to the user's activity. Thus, in this case the MAC matching is performed to obtain better identification results. If there is no matching between the domain names requested by the single host and the domain names stored in the database, then the BDM considers this as a new domain name which is not stored in its database. In this case, the BDM performs MAC matching between the MAC address of this host and the blacklisted MAC addresses stored in the database and checks if this host sends repeated query to this domain name as depicted in Figure 4. However, if there is no matching, then the BDM considers it as a normal domain name, and stores it in its database.

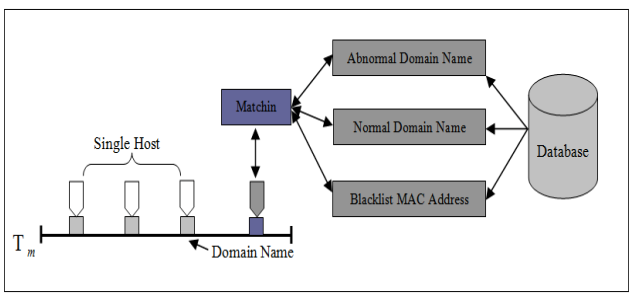

Figure 4.  Matching Single Host with Database

If there is a matching (*the host activity sends repeated queries*), then the BDM sends alarm to the network administrator to block this new abnormal domain name, because this behavior caused by the bots in the single hosts (*i.e. early infection stages*). However, this new domain name is stored in database as abnormal domain name, because it is a new domain name that the botmaster will use to communicate with his bots and issue further commands in the future. This allows BDM to predict the new attackers' domain name and send alarm to the network administrator to block it before the group of infected hosts appear and acquire it, which prevents the botnet activity on the network.

## VIII.  VALIDATION

Test is performed on BDM at NAv6[1] Network in USM to capture the real DNS requests from the hosts. The iNetmon[2] project was utilized in this test. The BDM runs on Intel core2 Duo 2.00 GHz CPU and 2.00 GB memory with Microsoft Windows Vista operating system. The simulator BotDNS is installed and runs in different hosts of the NAv6 network and set up requests a specific domain name (i.e. *www.xxx.com*) periodically, which is every 60 seconds.

The action taken in this scenario is to store this domain name as an abnormal domain name and send alarm to the network administrator to block it before the infected hosts request for this new domain name and consequently prevent the network from botnet activity in the future.

### A.  Performance Test Results

The experiment was carried to capture the real DNS request from the hosts in NAv6 network. The obtained results of classifying domain names are stored in the BDM database. These results contain more than 2000 domain names, which are requested by the hosts during the experiment.

The classification of domain names into normal domain names caused by the legitimate hosts and abnormal domain names caused by Botnet activity is shown in Figure 6. The threshold value for the Botnet domain name is set within 0.8 to 1 and for legitimate domain name, it is set within 0 to less than 0.8, based on the Jaccard similarity value as mentioned earlier in Table 2.

---

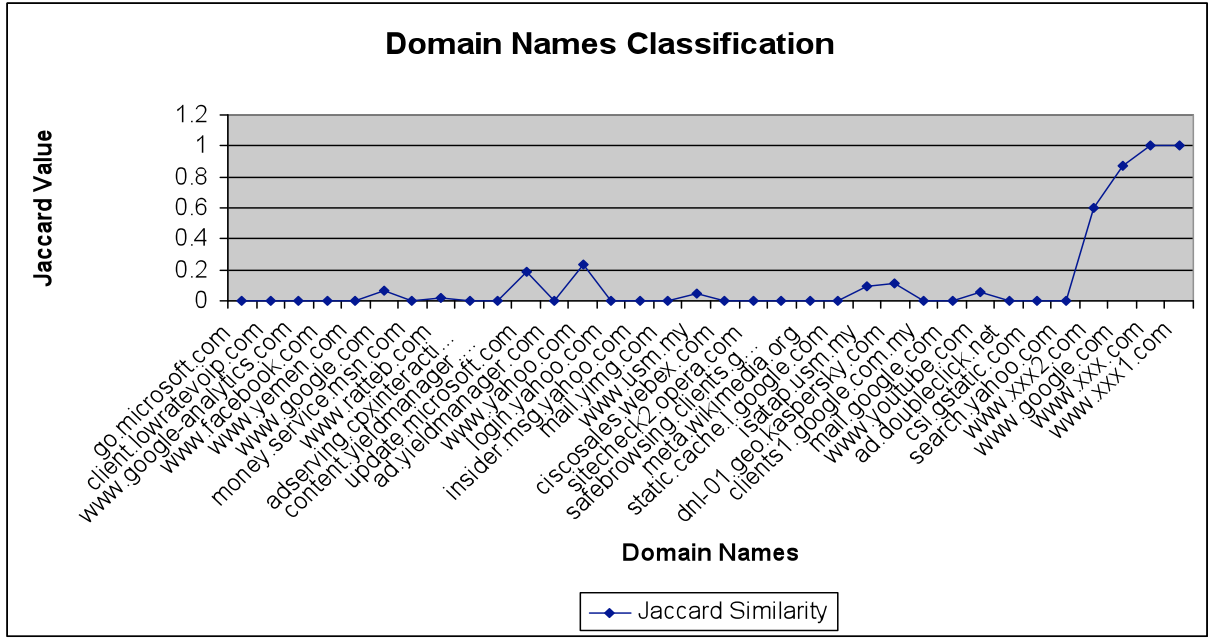

Figure 5.   Domain Names Classification Based on Jaccard Similarity

## B. False Positive

A false positive classifies a normal domain name as abnormal domain name. The false positive rate $F_{pr}$ can be calculated with Equation 3 [8]:

$$F_{pr} = \frac{F_P}{T_N} \qquad (3)$$

Where, $F_P$ refers to the number of false positive domain names detected and $T_N$ refers to the total number of true positive, $T_P$ domain name detected and the number of false positive, $F_P$ domain names detected.

We repeated the above test three times[*]; hence, it is observed that in the first experiment there is no false positive generated, whereas most legitimate domain names are classified correctly as normal domain names, so the false positive rate is 0% for this experiment.  In the second experiment, there is a normal domain that is classified as abnormal domains, so the false positive is present in this experiment, where the BDM classifies Google as abnormal domain name. During the second experiment period, there are three abnormal domain names detected and one legitimate domain name classified as abnormal domain name. The rate of false positive generated during this experiment is 33%.

$$F_{Pr} = \frac{1}{3} \approx 33\%$$

Finally the third experiment is same as the first experiment where no false positive is generated, so the false positive rate is 0% in this experiment. By taking the average value for the false positive rate from these three experiments, the rate of false positive generated during the experiments is 11%.

$$\text{Average } F_{Pr} = \frac{0 + 33 + 0}{3} \approx 11\%$$

---

[*] These tests were repeated three times for average reading. While the detection experiment was performed over one week.

From the ratio of false positive rate, the detection rate of BDM can be obtained. The detection rate, $D_r$ can be calculated with Equation 4 [8]:

$$D_r = \frac{T_P}{T_N} \qquad (4)$$

From the first and third experiments, there is no false positive found, hence the detection rate is approximately 100%. However, in the second experiment the detection rate, $D_r$ is approximately 67%:

$$D_r = \frac{2}{3} \approx 67\%$$

By taking the average of detection rate, it can be observed that BDM has average detection rate of 89% during these three experiments that are considered acceptable.

$$\text{Average } D_r = \frac{100 + 67 + 100}{3} \approx 89\%$$

## IX. CONCLUSION

In this paper, a simple framework is proposed called the BDM for botnet detection in a network environment. The framework consists of three phases that capture the DNS traffic, extract the MAC address, and query name from this DNS packet and store it in the database for further analysis. After that, the BDM classifies the DNS traffic that is issued by blocks of hosts (group behavior) and single host (individual behavior).

The proposed method depends on monitoring the DNS traffic and exploiting the behavior of Botnet. The Botnet is detected in blocks of hosts (group behavior) by measuring the degree of similarities between any two blocks of hosts requesting the same domain name at different time intervals based on the Jaccard similarity coefficient $S_j$. The MAC address is used as host's identifier instead of IP address.

The results of the experiments on the NAv6 network shows that the BDM is robust and works well, with the average detection rate of about 89%. It is capable of classifying the domain names into normal and abnormal domain names, consequently, detecting the Botnet activity within the network. The BDM classifies domain names based on the Jaccard similarity value. However, during the experiments, in average the BDM generated false positive and false negative rate, which was approximately 11% in each case.

The main limitation in the testing is that the average detection rate is based on three experiments only. A future work could be to do more experiments for better accuracy. Another future work is considering the improvement of the BDM to enable identifying and tracing back the infected hosts within the monitored network as well as enhancing and increasing the detection rate by minimizing the false positive/negative alerts by incorporating different statistical methods such as chai-square along with the jaccard similarities coefficient.